\documentclass[prd,showkeys,floatfix,nofootinbib,
               preprint,12pt,
               tightenlines,fleqn]{revtex4} 

\usepackage{amsmath,amssymb,revsymb,graphicx,dcolumn}
\usepackage{leftidx}

\newcommand{\beq}{\begin{equation}}
\newcommand{\eeq}{\end{equation}}
\newcommand{\beqa}{\begin{eqnarray}}
\newcommand{\eeqa}{\end{eqnarray}}
\newcommand{\bsubeqs}{\begin{subequations}}
\newcommand{\esubeqs}{\end{subequations}}

\begin{document}
\title[De-Sitter-spacetime instability]
      {Freely-moving observer in (quasi) anti de Sitter space\vspace*{5mm}}
\author{Slava Emelyanov}
\email{viacheslav.emelyanov@physik.uni-muenchen.de}
\affiliation{Arnold Sommerfeld Center for
Theoretical Physics,\\
Ludwig Maximilian University (LMU),\\
80333 Munich, Germany\\}

\begin{abstract}
\vspace*{2.5mm}\noindent
A quantum scalar field in anti de Sitter space is considered in 
two coordinate systems: static and FRW-like ones. It is shown that quantum vacua corresponding to each 
of these coordinatizations are not unitary equivalent. A choice of a physical ground state between these 
vacua is discussed under different setups.
\end{abstract}


\maketitle

\section{Introduction}
\label{sec:Introduction}

Historically, Minkowski spacetime is a manifold wherein a quantization of a field was first performed. The Poincar\'{e}-invariant 
no-particle state is known as Minkowski vacuum that is associated with inertial observers. 40 years ago it was 
discovered that a quantum vacuum defined by a uniformly accelerated observer is not unitary 
equivalent to Minkowski one~\cite{Fulling}. Moreover, Minkowski vacuum for such an observer is seen as a thermal 
bath of particles defined with respect to the observer's reference frame. This phenomenon is known as Davies-Unruh 
effect~\cite{Davies,Unruh,Birrell&Davies,Mukhanov&Winitzki,Crispino&Higuchi&Matsas}. Later it was also shown that the 
notion of a particle of a massive 
field in Milne spacetime is inequivalent to the Minkowski particle, where Milne vacuum has been defined as a no-particle 
state that becomes the conformal vacuum in the massless limit~\cite{Sommerfield}. If the conformal quantum field occupies 
Milne vacuum, then a radiation with a Planck spectrum and negative energy density is present in it~\cite{Birrell&Davies}.

Another example of a manifold where similar situation takes place is de Sitter space. De Sitter hyperboloid is a curved 
maximally symmetric spacetime with a negative, constant curvature $R = - 12H_\text{dS}^2$. It can be covered entirely or 
partially by various coordinate systems~\cite{Mukhanov,Griffiths&Podolsky}. Quantization of a field performed in different 
coordinatizations 
generally leads to a definition of different quantum vacua. In particular, a comoving observer in dS defines its own no-particle 
state $|0_\text{S}\rangle$ that is associated with the static coordinate frame, where the observer is at its origin.  
The Unruh-DeWitt detector moving along a time-like geodesic in the de Sitter hyperboloid will 
be excited like it is in the thermal bath with the Hawking-Gibbons temperature $T_\text{HG} = H_\text{dS}/2\pi$~\cite{Gibbons&Hawking}.
Thus, the vacuum state associated with the field quantized in static dS is not unitary equivalent to the closed one, 
i.e. $|0_\text{S}\rangle \nsim |0_\text{CT}\rangle$, where $|0_\text{CT}\rangle$ is the Chernikov-Tagirov ground 
state, as well as to the flat vacuum~\cite{Birrell&Davies}.

In the present paper we shall consider anti de Sitter hyperboloid that is also a curved maximally symmetric manifold, but with a
positive, constant curvature. There are known variety of coordinate patches which parameterize the whole or some 
set of the AdS hyperboloid points~\cite{Griffiths&Podolsky}. We shall take into our consideration two of them, namely static 
(or global) 
and open (or FRW-like with the negative spatial curvature, see~\cite{Griffiths&Podolsky}) coordinates. Static coordinates cover
the entire AdS, whereas open coordinates embrace merely half of it. Therefore one may a priori await that quantization 
performed in each coordinate systems leads to inequivalent quantum vacua. Thus, the aim of our research is to uncover how 
open vacuum $|0_\text{O}\rangle$ is related with static one $|0_\text{S}\rangle$ (both defined below) as  well as to figure out
whether a freely-moving detector in AdS will be somehow excited.

In Section \ref{sec:ads4}, a real scalar field model is considered in four-dimensional AdS. 
For the sake of avoiding unnecessary complications, we assume that it is invariant under 
the conformal symmetry.
The question of a physical vacuum state is examined by analysing cosmological consequences resulting from taking into account quantum corrections at one-loop approximation in semiclassical Einstein equation for each of them.

In Section \ref{sec:ads2}, a two-dimensional massive scalar field is studied. In this case, we shall further investigate 
the vacua under different setups as well as compare them with the Minkowski state.

We work throughout in units implying $\hbar = c = k_\text{B} = G = 1$. The signature of the metric tensor is $(+,-,-,-)$ and
$R_{\mu\nu} = \partial_{\lambda}\Gamma_{\mu\nu}^{\lambda} - \partial_{\nu}\Gamma_{\mu\lambda}^{\lambda}
+ \Gamma_{\rho\lambda}^{\lambda}\Gamma_{\mu\nu}^{\rho} - \Gamma_{\rho\mu}^{\lambda}\Gamma_{\lambda\nu}^{\rho}$.

\section{Conformal scalar field in AdS$_4$}
\label{sec:ads4}

Anti de Sitter spacetime 
can be represented as a four dimensional hyperboloid  
\beqa\label{ads4:hyperboloid}
x_0^2 - x_1^2 - x_2^2 - x_3^2 + x_4^2 &=& H^{-2}
\eeqa
embedded in $\mathbf{R}^5$ with the line element $ds^2 = dx_0^2 - dx_1^2 - dx_2^2 - dx_3^3 + dx_4^2$~\cite{Hawking&Ellis}. 

Anti de Sitter space is not globally hyperbolic. Its topology is given by $\mathbf{S}\times\mathbf{R}^3$, where the circle 
$\mathbf{S}$ corresponds to time coordinate. This means it contains closed time-like curves. A standard approach for avoiding casual 
paradoxes in AdS is to unwrap  $\mathbf{S}$ and to consider its  universal covering $\mathbf{R}$~\cite{Hawking&Ellis}. In the following, we deal with universal covering space only. Besides, Cauchy surface is absent in it. The quantization procedure is performed by imposing certain boundary conditions at spatial time-like infinity which ensure, in particular, the energy conservation~\cite{Avis&Isham&Storey,Breitenlohner&Freedman}.

As has been above mentioned, we shall deal with static and open coordinates (see App.~\ref{app:coordinates}). 
Static coordinates cover the entire hyperboloid and, as a consequence, one needs to impose the boundary conditions on the 
field to have a well-defined quantum theory~\cite{Avis&Isham&Storey,Breitenlohner&Freedman}. Open coordinates cover merely a part of the hyperboloid that is globally hyperbolic. 

\paragraph*{Static AdS$_4$} 

The mode expansions of the quantum scalar field conformally coupled with gravity in static coordinates read
\beqa
\hat{\phi}(x) &=& 
\sum\limits_{nlm}
\left\{
\begin{array}{l}
\hat{a}_{nlm}^{\phantom{}}u_{nlm}^{\phantom{\dagger}}(x) + \hat{a}_{nlm}^{\dagger}u_{nlm}^*(x)\,,
\\[2.5mm]
\hat{b}_{nlm}^{\phantom{}}v_{nlm}^{\phantom{\dagger}}(x) + \hat{b}_{nlm}^{\dagger}v_{nlm}^*(x)\,,
\end{array}
\right.
\eeqa
where star denotes the complex conjugation.
The positive-frequency modes $u_{nlm}(x)$ and $v_{nlm}(x)$ with respect to the time translation operator are
\bsubeqs\label{eq:ads4:modes}
\beqa
u_{nlm}(x) &=& A_{nl}\,e^{-i(2n+2+l)\eta}\,Y_{lm}(\theta,\varphi)(\sin\chi)^l\cos\chi\,C_{2n+1}^{l+1}(\cos\chi)\,,
\\[2.5mm]
v_{nlm}(x) &=& B_{nl}\,e^{-i(2n+1+l)\eta}\,Y_{lm}(\theta,\varphi)(\sin\chi)^l\cos\chi\,C_{2n}^{l+1}(\cos\chi)\,,
\eeqa
\esubeqs
where $n,l \in \mathbb{N}_0$ and $m \in \{-l,...,+l\}$, $A_{nl}$ and $B_{nl}$ are normalization 
coefficients~\cite{Avis&Isham&Storey,Breitenlohner&Freedman}.
The static vacuum $|0_\text{S}\rangle = |0_{\text{S}_1}\rangle\otimes|0_{\text{S}_2}\rangle$ is annihilated by both $\hat{a}_{nlm}$ 
and  $\hat{b}_{nlm}$: $\hat{a}_{nlm}|0_{\text{S}_1}\rangle = \hat{b}_{nlm}|0_{\text{S}_2}\rangle = 0$. The Fock space of particle states is built by cyclic operations of the creation operators $\hat{a}_{nlm}^\dagger$ and $\hat{b}_{nlm}^\dagger$ on $|0_\text{S}\rangle$.

\paragraph*{Open AdS$_4$} 

The mode expansion of the field in open coordinates is given by
\beqa\label{eq:ads2:modes:open}
\hat{\phi}(\bar{x}) &=& \sum\limits_{lm}\int\limits_0^{+\infty}d\omega\Big(
\hat{c}_{\omega lm}^{\phantom{\dagger}}w_{\omega lm}^{\phantom{\dagger}}(\bar{x}) 
+ \hat{c}_{\omega lm}^{\dagger}w_{\omega lm}^*(\bar{x})\Big)\,,
\eeqa
where the mode functions $w_{\omega lm}(\bar{x})$ regular at $\bar{\chi} = 0$ are
\beqa\label{eq:ads4:w-modes}
w_{\omega lm}(\bar{x}) &=& 
\frac{C_{\omega l}}{a(\bar{\eta})}\,e^{-i\omega\bar{\eta}}\,Y_{lm}(\bar{\theta},\bar{\phi})
\left(\sinh\bar{\chi}\right)^{-\frac{1}{2}}P_{i\omega-1/2}^{-l-1/2}(\cosh\bar{\chi})\,,
\eeqa
where $C_{\omega l}$ is a real normalization coefficient~\cite{Sasaki&Tanaka&Yamamoto}. 
Open vacuum $|0_\text{O}\rangle$ is defined as being annihilated by  $\hat{c}_{\omega lm}$. The Fock space is 
generated by acting by the creation operator $\hat{c}_{\omega lm}^\dagger$ on this vacuum.

In contrast to the static case, there is no time-like Killing vector in open AdS that can be used to
define the Hamiltonian operator being the representation of the time translation on space of quantum states. However, the conformal 
time-like Killing vector $K = \partial_{\bar{\eta}}$ allows us to introduce
\beqa\label{eq:ads4:h}
\hat{H} &=& \int d\Sigma_{\mu}\hat{T}^{\mu}_{\nu}K^{\nu}
\;\;=\;\; \frac{1}{2}\sum\int d\omega \omega \Big(\hat{c}_{\omega lm}^{\dagger}\hat{c}_{\omega lm}^{\phantom{\dagger}}
+ \hat{c}_{\omega lm}^{\phantom{\dagger}}\hat{c}_{\omega lm}^{\dagger}\Big)\,,
\eeqa
where $T_{\mu\nu}$ is the energy-momentum tensor of the field. The operator $\hat{H}$ is conserved due to the conformal 
symmetry. It can be interpreted as a generator of the time translation acting on the rescaled field 
$\hat{\phi}_\text{r}(\bar{x}) \equiv a(\bar{\eta})\hat{\phi}(\bar{x})$, i.e. $\mathcal{L}_{K}\hat{\phi}_\text{r} 
= i\big[\hat{H},\hat{\phi}_\text{r}\big]$, 
where $\mathcal{L}_{K}$ is the Lie derivative along $K$. Thus, the rescaled modes \eqref{eq:ads4:w-modes} are positive 
frequency modes with respect to $\partial_{\bar{\eta}}$ and $|0_\text{O}\rangle$ is called the conformal 
vacuum~\cite{Birrell&Davies}.

Although the conformal time $\bar{\eta}$ lies on the whole real line $\mathbf{R}$, the expansion \eqref{eq:ads2:modes:open} of the 
quantum field is performed in terms of the physical time $\bar{t}$ only for $(0,\pi/H)$ that corresponds to the upper wedge depicted in 
Fig.~\ref{fig:embeddings}.
However, the same expansion can be used for the lower wedge, wherein $\bar{t} \in (-\pi/H,0)$. 
Shortly, a difference between these wedges will be discussed. The modes in the rest of the wedges can be straightforwardly 
obtained from the upper and lower ones.

\subsection{Bogolyubov transformation}

A curve $\chi = 0$ and $\Delta\theta =\Delta\phi = 0$ is a geodesic in static AdS. 
An observer that moves along it will detect nothing, provided the quantum field is in $|0_{\text{S}_a}\rangle$, where 
$a \in \{1,2\}$. Indeed, the response function $\mathcal{F}(E)$ of the Unruh-DeWitt detector with respect to $|0_{\text{S}_a}\rangle$ is 
\beqa
\mathcal{F}(E) &\propto& \int\limits_{-\infty}^{+\infty}d\Delta\tau\,e^{-iE\Delta\tau}\,
\left(\frac{1}{\sin^2\left(\frac{1}{2}H\Delta\tau - i\varepsilon\right)} - \frac{(-1)^a}{\cos^2\left(\frac{1}{2}H\Delta\tau - i\varepsilon\right)}\right)\,,
\quad \varepsilon \rightarrow +0\,,
\eeqa
where $E$ is an eigenstate of detector's Hamiltonian, $\tau$ the proper time along observer's geodesic~\cite{Birrell&Davies}.
Representing the integrand as a series (1.422.2 and 1.422.4 in~\cite{Gradshteyn&Ryzhik}) and regarding it as a contour one over 
a closed contour in an infinite semicircle in a lower-half $\Delta\tau$ plane, one obtains $\mathcal{F}(E) = 0$ that
proves the statement.

A curve $\Delta\bar{\chi} = 0$ and $\Delta\theta =\Delta\phi = 0$ represents a comoving geodesic in open AdS. These geodesics 
belong to a time-like set of integral curves of the vector $\xi = \sin\eta\cos\chi\,\partial_{\eta} + \cos\eta\sin\chi\,\partial_{\chi}$. This vector 
field is time-like in open AdS, but is future-directed for $\eta \in (2\pi k, \pi + 2\pi k)$ and past-directed for $\eta \in (-\pi +2\pi k,2\pi k)$, 
where $k \in \mathbb{N}$. Outside of open AdS the vector field $\xi$ is space-like and null on the horizons  $\eta = \pm\chi +\pi k$. 
The vacuum state invariant under $\exp(-i\hat{H}\bar{\eta})$ is the conformal one, i.e.  $|0_\text{O}\rangle$. 
Since for any comoving observer a relation between the conformal time $\bar{\eta}$ and the proper one $\tau$ does not 
depend on spatial coordinates, the Unruh-DeWitt detector will be similarly exited or unexcited for any of them. 

At the points $\chi = \bar{\chi} = 0$ and $\theta,\phi = \text{const}$, two vacua are defined. As has been shown above, the detector
moving along this geodesic will remain unexcited if the field in $|0_{\text{S}_a}\rangle$. However, this detector may response if the field is
in $|0_\text{O}\rangle$. In order to find how these vacua are related to each other, we perform the Bogolyubov transformation that represents 
the annihilation $\hat{c}_{\omega lm}^{\phantom{}}$ and creation $\hat{c}_{\omega lm}^\dagger$ operators defining conformal vacuum 
$|0_\text{O}\rangle$ as a linear combinations of analogous operators associated with $|0_{\text{S}_a}\rangle$. Note, there is no inverse 
Bogolyubov transformation, because open coordinates do not cover the entire hyperboloid. 

This canonical, unitary map is completely determined by the Bogolyubov coefficients
\begin{equation}\label{eq:ads4:bogolyubov-coefficients}
\begin{array}{lcl}
\alpha_{\omega lm, nl^{\prime}m^{\prime}} &=& \big(u_{nl^{\prime}m^{\prime}}(x),w_{\omega lm}(\bar{x})\big)\,,
\\[2.5mm]
\beta_{\omega lm, nl^{\prime}m^{\prime}} &=& \big(u_{nl^{\prime}m^{\prime}}^*(x),w_{\omega lm}(\bar{x})\big)\,,
\end{array}
\end{equation}
where the round brackets denote the Klein-Gordon scalar product~\cite{Birrell&Davies}. We have used in 
\eqref{eq:ads4:bogolyubov-coefficients} 
an expansion of the quantum field expressed through $u$-modes for the sake of concreteness. The same result can be shown is
valid for the expansion through $v$-modes. Below
we discuss the upper wedge. At the end of this subsection we shall point out which changes must be made to accommodate
\eqref{eq:ads4:bogolyubov-coefficients} with the lower wedge.

Calculation of the coefficients is considerably simplified if one takes  a hypersurface of equal-time events $\Sigma$ specified by $\bar{\eta} = 0$. 
Having chosen it, one obtains
\begin{equation}\label{eq:ads4:bogolyubov-coefficients-1}
\begin{array}{lcl}
\alpha_{\omega lm, nl^{\prime}m^{\prime}} &=& N_{ln}(\omega)\,\delta_{ll^{\prime}}\,\delta_{mm^{\prime}}\,M_{n,l}^{+}(\omega)\,,
\\[2.5mm]
\beta_{\omega lm, nl^{\prime}m^{\prime}} &=& N_{ln}(\omega)\,\delta_{ll^{\prime}}\,\delta_{mm^{\prime}}\,M_{n,l}^{-}(\omega)\,,
\end{array}
\end{equation}
where $N_{nl}(\omega)$ is written down in App. D and
\beqa\label{eq:ads4:m}
M_{n,l}^{\pm}(\omega) &\equiv& \int\limits_0^{+\infty}d\bar{\chi}\,
\left(\omega \pm \frac{2 + l + 2n}{\cosh\bar{\chi}}\right)\frac{(\sinh\bar{\chi})^{l+\frac{3}{2}}}{(\cosh\bar{\chi})^{l+2}}\,
\big(\tanh\left(\bar{\chi}/2\right)\big)^{l+\frac{1}{2}}
\\[0.5mm]\nonumber
&\times&
_2F_1\left(-n,n+l+2;\frac{3}{2};\frac{1}{\cosh^2(\bar{\chi})}\right)
{}_2F_1\left(\frac{1}{2}-i\omega,\frac{1}{2}+i\omega;l+\frac{3}{2};-\sinh^2\left(\bar{\chi}/2\right)\right)\,,
\eeqa
where the Gegenbauer $C_{\nu}^{\mu}(z)$ and the associated Legendre $P_{\nu}^{\mu}(z)$ functions have been rewritten via 
the hypergeometric one $_2F_1(\alpha,\beta;\gamma;z)$~\cite{Gradshteyn&Ryzhik}.

The integrals \eqref{eq:ads4:m} for $l=0$ can be exactly evaluated and are
\beqa
M_{n,0}^{\pm} &=&\mp
\frac{\pi^2\exp\left(\pm\frac{\pi\omega}{2}\right)}{\sqrt{2}\sinh^2(\pi\omega)}\,
\frac{{}_2F_1\left(-2-2n,i\omega;i\omega-1-2n;-1\right)}{\Gamma\left(2n+3\right)\Gamma\left(1-i\omega\right)\Gamma\left(i\omega-1-2n\right)}\,,
\eeqa
where $J_n^{\pm}(\omega)$ and $I_n^{\pm}(\omega)$ given in App.~\ref{app:integrals} have been used. Thus, we obtain
\beqa
\alpha_{\omega 00, n00} &=& -e^{\pi\omega}\beta_{\omega 00, n00}\,.
\eeqa
For arbitrary $l \ge 0$, one can express $M_{n,l+1}^{\pm}$ through $M_{n,l}^{\pm}$ and $M_{n+1,l}^{\pm}$ as follows
\beqa\label{eq:ads4:mp}
M_{n,l+1}^{\pm} &=& \frac{2l+3}{4\big(\omega^2 + (l+1)^2\big)}\Big((3+2l+2n)M_{n,l}^{\pm} - (2n+3)M_{n+1,l}^{\pm}\Big)\,
\eeqa
(see App. \ref{app:property} for details). Consequently, one has
\beqa\label{eq:ads4:ratio}
\alpha_{\omega lm, n^{\prime}l^{\prime}m^{\prime}} &=& -e^{\pi\omega}\beta_{\omega lm, n^{\prime}l^{\prime}m^{\prime}}\,.
\eeqa

Repeating the same kind of calculations for the Bogolyubov coefficients in the lower wedge, 
one finds that they are equal to \eqref{eq:ads4:bogolyubov-coefficients-1} multiplied by $(-1)^{n+1}$, where $n$ is odd
for $u$-modes, and even for $v$-modes.

\subsection{Relation between static $|0_{\text{S}}\rangle$ and open $|0_\text{O}\rangle$ vacua}

Vacua $|0_{\text{S}_a}\rangle$ and $|0_\text{O}\rangle$ are not unitary equivalent.
Despite of their orthogonality, no-particle state $|0_\text{O}\rangle$ can formally 
be represented as a state realized by infinite number of particles defined with respect to $|0_{\text{S}_a}\rangle$.
In order to see this, one has to represent the annihilation operator $\hat{c}_{\omega lm}$ as follows\footnote{In the following
we shall discuss only the case $a = 1$. The same result can be obtained for $a=2$ in the analogous manner.}
\beqa
\hat{c}_{\omega lm}^{\phantom{}} &=& \cosh\theta_{\omega}\,\hat{d}_{\omega lm}^{\phantom{}} -  \sinh\theta_{\omega}\,\hat{f}_{\omega lm}^\dagger\,,
\eeqa
where by definition
\beqa\label{eq:ads4:d-f-operators}
\hat{d}_{\omega lm}^{\phantom{}} &\equiv& \frac{1}{\cosh\theta_{\omega}}\,\sum\limits_{n^{\prime}l^{\prime}m^{\prime}}
\alpha_{\omega lm, nl^{\prime}m^{\prime}}\hat{a}_{n^{\prime}l^{\prime}m^{\prime}}\,, \quad
\hat{f}_{\omega lm}^{\phantom{}} \;\;\equiv\;\; \frac{-1}{\sinh\theta_{\omega}^*}\,\sum\limits_{n^{\prime}l^{\prime}m^{\prime}}
\beta_{\omega lm, nl^{\prime}m^{\prime}}^*\hat{a}_{n^{\prime}l^{\prime}m^{\prime}}\,.
\eeqa
It is shown in App.~\ref{app:relations} that if one takes $\tanh\theta_{\omega} = e^{-\pi\omega}$, then
\bsubeqs
\beqa\label{eq:ads4:commutations1}
\big[\hat{d}_{\omega lm}^{\phantom{}},\hat{d}_{\omega^{\prime}l^{\prime}m^{\prime}}^{\dagger}\big] &=&
\big[\hat{f}_{\omega lm}^{\phantom{}},\hat{f}_{\omega^{\prime}l^{\prime}m^{\prime}}^{\dagger}\big] \;\;=\;\;
\delta(\omega - \omega^{\prime})\,\delta_{ll^{\prime}}\,\delta_{m m^{\prime}}\,,
\\[1.5mm]\label{eq:ads4:commutations2}
\big[\hat{d}_{\omega lm}^{\phantom{}},\hat{f}_{\omega^{\prime}l^{\prime}m^{\prime}}^{\phantom{}}\big] &=&
\big[\hat{d}_{\omega lm}^{\phantom{}},\hat{f}_{\omega^{\prime}l^{\prime}m^{\prime}}^{\dagger}\big] \;\;=\;\; 0
\eeqa
\esubeqs
hold. Such a pair of operators can be used to obtain
\beqa\label{eq:ads4:relation}
|0_\text{O}\rangle &=& Z^{-1}\exp\left(\sum\int d\omega e^{-\pi\omega}
\hat{d}_{\omega lm}^{\dagger}\hat{f}_{\omega lm}^{\dagger}\right)|0_{\text{S}_1}\rangle\,, \quad
\ln(Z) \;=\; \frac{\pi}{24}\delta(0)\sum(2l+1)\,,
\eeqa
where summations are implied over the discrete indices~\cite{Umezawa}. 
Both operators $\hat{d}_{\omega lm}$ and $\hat{f}_{\omega lm}$ annihilate static vacuum $|0_{\text{S}_1}\rangle$. 
Since  $\hat{d}_{\omega lm}|0_\text{O}\rangle = e^{-\pi\omega}\hat{f}_{\omega lm}^{\dagger}|0_\text{O}\rangle$
and $\hat{f}_{\omega lm}|0_\text{O}\rangle = e^{-\pi\omega}\hat{d}_{\omega lm}^{\dagger}|0_\text{O}\rangle$, one has
\beqa\label{eq:ads4:ovevd}
\langle 0_\text{O}|\hat{d}_{\omega lm}^{\dagger}\hat{d}_{\omega^{\prime}l^{\prime}m^{\prime}}^{\phantom{}}|0_\text{O}\rangle
&=& \frac{1}{e^{2\pi\omega} -1}\,\delta(\omega - \omega^{\prime})\delta_{ll^{\prime}}\delta_{mm^{\prime}}
\eeqa
and the same for the $f$-operators. The equation \eqref{eq:ads4:ovevd} implies that the conformal vacuum $|0_\text{O}\rangle$ 
can be regarded as a thermal bath of particles created by applying 
$\hat{d}_{\omega lm}^{\dagger}$ on $|0_{\text{S}_1}\rangle$.

The operators \eqref{eq:ads4:d-f-operators} are not independent: $\hat{f}_{\omega lm} = i(-1)^l\,\hat{d}_{-\omega lm}$.
There must be a unitary equivalent quantization of the field in static AdS, such that the rescaled field is 
expanded through modes which are eigenfunctions of the vector field $\xi$.
The creation operator must be given then by $\hat{d}_{p lm}^{\dagger}$, whereas $p \in \mathbf{R}$ and $\omega = |p|$. 
Except for the fact that vector $\xi$ is not one of the generators of the anti de Sitter group $\text{SO}(3,2)$, this resembles 
an equivalent quantization in Minkowski space, wherein the field can also be expanded through modes being eigenfunctions of 
the Lorentz operator~\cite{Crispino&Higuchi&Matsas}.

Thus, a freely-moving detector will be thermally excited, provided the quantum field is in the conformal vacuum. 
Expressing $\omega$ via the physical frequency, i.e. $\omega_\text{ph} = \omega/a(\bar{t})$,  one obtains that
the temperature ascribed to the thermal condensate is equal to $T_\text{AdS} = (2\pi a(\bar{t}))^{-1}$, where 
$a(\bar{t}) = H^{-1}\cos(H\bar{t})$. At this point, we merely note that $T_\text{AdS}$ diverges when $\cos(H\bar{t}) = 0$.
Below we shall discuss a physical consequence of this result.

\subsection{Choice of physical vacuum}

In order to address a question of physicality of the vacua, we shall compute the renormalized energy-momentum tensor of 
the field in each of them, because the quantum field under consideration reveals itself physically through gravity. For instance,
the renormalized stress tensor for the conformal scalar field in static vacuum in dS
diverges at the event horizon, while it is finite and proportional to the metric for closed as well as flat 
vacua~\cite{Candelas&Dowker2902}. 
Thus, the physical vacuum occupied by the quantum field in de Sitter space corresponds to Chernikov-Tagirov
state~\cite{Sciama&Candelas&Deutsch}. 
\begin{figure}
\includegraphics[width=4.95cm]{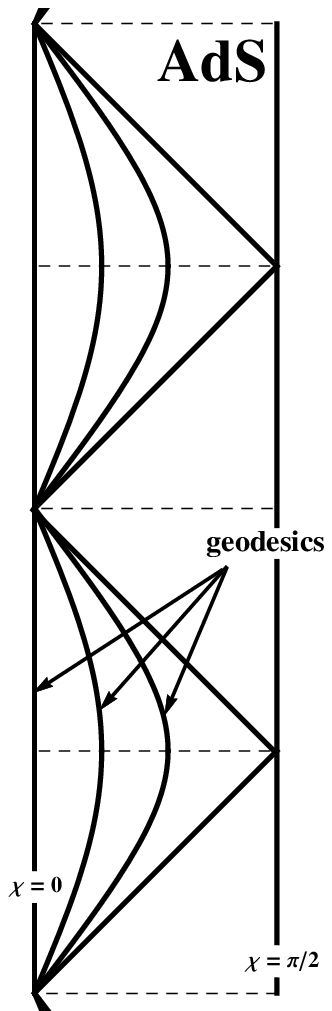}\quad\quad
\includegraphics[width=4.95cm]{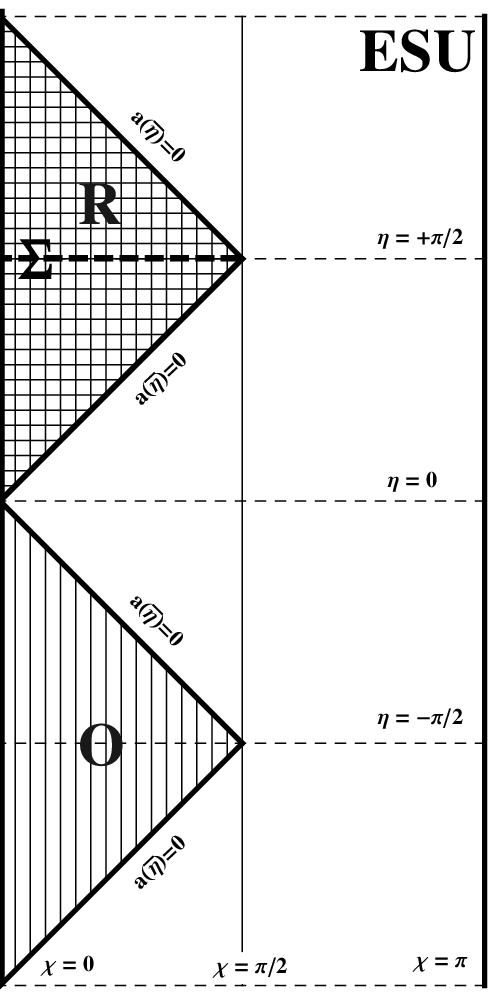}
\caption{Left: conformal diagram of Anti de Sitter space. Right: conformal diagram of Einstein static universe (ESU) with
embedded open AdS (O wedges filled by vertical lines) and Rindler (R wedge filled by horizontal lines) spaces in it.}
\label{fig:embeddings}
\end{figure}

The Euclidean section of entire AdS$_4$ is the hyperbolic space H$^4$. The $\zeta$-function regularization method 
has been applied in H$^4$ to derive the exact form of the one-loop effective action~\cite{Camporesi&Higuchi}. 
In this approach, the renormalized stress tensor corresponds to $|0_{\text{S}_1}\rangle$, as it follows from the
comparison of two-point correlation functions calculated by $\zeta$-function method and directly summing the 
modes~\cite{Burgess&Luetken}. Thus, one has
\beqa\label{eq:ads4:ren:1}
\langle \hat{T}_{\nu}^{\mu} \rangle_{\text{S}_1} &=& \frac{H^4}{960\pi^2}\,\delta_{\nu}^{\mu}\,,
\eeqa
where $\langle ...\rangle_{\text{S}_1}$ denotes the vacuum expectation value with respect to $|0_{\text{S}_1}\rangle$, i.e.  
$\langle 0_{\text{S}_1}|...|0_{\text{S}_1}\rangle$. 

This result can be obtained in a different way. The Feynman two-point function $G_\text{F}^1(x,y)$ corresponding to
$|0_{\text{S}_1}\rangle$ satisfies the scalar field equation with two $\delta$-sources which are located at $y$ and its antipodal
point $y_\text{A} = -y$~\cite{Avis&Isham&Storey}. The effective action is
\beqa
W[\,g_{\mu\nu}] &=& -\frac{i}{2}\ln\det\big(-\hat{G}_\text{F}(1+\hat{\pi})\big)\,,
\eeqa
where $\langle x|\hat{G}_\text{F}|y\rangle = G_\text{F}(x,y)$ satisfies the field equation with the standard $\delta$-source 
and $\hat{\pi}$ is the inversion operator: $\hat{\pi}|x\rangle = |-x\rangle$~\cite{Birrell&Davies,DeWitt}. Since the determinant of a matrix product equals the product of their determinants, one
can get rid of $\ln\det(1+\hat{\pi})$ by an infinite shift of the effective action $W[\,g_{\mu\nu}]$ that does not depend on the metric and, 
hence, has no physical consequences. Then, following the ordinary procedure, the trace anomaly can be derived that gives 
\eqref{eq:ads4:ren:1}. Hence, one obtains 
\beqa\label{eq:ads4:ren:2}
\langle \hat{T}_{\nu}^{\mu} \rangle_{\text{S}_2} &=& \frac{H^4}{960\pi^2}\,\delta_{\nu}^{\mu}\,.
\eeqa

This derivation may be supported by the following independent result derived by the point-splitting method: the renormalized stress 
tensor of a conformal scalar field that inhabits in half Einstein static universe (topologically $\mathbf{R}\times\mathbf{S}^3$) and satisfies Dirichlet 
or Neumann boundary conditions at the equator of the sphere $\mathbf{S}^3$ is exactly equal to the stress tensor of the field if it would 
be defined on entire $\mathbf{S}^3$~\cite{Kennedy&Unwin}.

To compute $\langle \hat{T}_{\nu}^{\mu} \rangle_\text{O}$, one needs to take into account 
that wedges of open AdS can be conformally mapped into Rindler spacetime (see Fig.~\ref{fig:embeddings}). Thus, one has
\beqa\label{eq:ads4:ren3}
\langle \hat{T}_{\nu}^{\mu} \rangle_\text{O} &=& \frac{H^4}{960\pi^2}\,\delta_{\nu}^{\mu}
- \frac{1}{480\pi^2a^4(\bar{\eta})}\left(\delta_{\nu}^{\mu} - \frac{4}{3}\,\delta_i^{\mu}\,\delta_{i\nu}\right)\,,
\eeqa
where summation with respect to $i$ running from 1 to 3 is implied~\cite{Candelas&Dowker2902}. 

The result \eqref{eq:ads4:ren3}  can be also derived by the point-splitting method supplemented by the symmetry properties of the problem. The anomalous trace specifies the renormalized
stress tensor of a conformal scalar field in conformally flat space up to the local, conserved, traceless tensor (denoted as $^{(4)}H_{\mu\nu}$ 
in~\cite{Birrell&Davies}). This non-geometric part of  $\langle \hat{T}_{\nu}^{\mu} \rangle_\text{O}$ is unambiguously fixed, if the renormalized 
stress tensor of the field in open static universe $\langle \hat{T}_{\nu}^{\mu} \rangle_\text{OSU}$ is known. The point-splitting 
approach gives $\langle \hat{T}_{\nu}^{\mu} \rangle_\text{OSU} = 0$~\cite{Bunch}.\footnote{Actually, it is sufficient to compute 
$\langle\hat{H}\rangle_\text{O}$ (see~\cite{Bunch}). If one takes the vacuum expectation value of $\hat{H}$ given in \eqref{eq:ads4:h} 
and renormalizes it by subtracting zero-point energy of the field quantized in Minkowski space in spherical coordinates, one then gets
$\langle\hat{H}\rangle_\text{O} = 0$.}  Taking then into account the isometries of the metric and the continuity equation, i.e. 
$\nabla_{\mu}\langle \hat{T}_{\nu}^{\mu} \rangle_\text{O} = 0$, one deduces 
\eqref{eq:ads4:ren3}.

The second term on the right-hand side of \eqref{eq:ads4:ren3} represents a radiation with a negative energy density
that vanishes in the classical limit: $\hbar \rightarrow 0$. It is worth noticing that the same situation occurs in Milne universe, 
wherein a comoving observer registers a thermal radiation with the negative energy density, provided the quantum field is in the conformal Milne vacuum~\cite{Birrell&Davies}. 

The renormalized stress tensor with respect to static vacuum $\langle \hat{T}_{\nu}^{\mu} \rangle_{\text{S}_a}$ behaves itself equally well throughout the space, while $\langle \hat{T}_{\nu}^{\mu} \rangle_\text{O}$ diverges at the horizons, where
$a(\bar{\eta})$ vanishes. It will be shown in the next subsection that AdS in open coordinates is unstable if the field occupies 
$|0_\text{O}\rangle$.

\subsection{Backreaction}

The Einstein equation with the quantum correction reads
\beqa\nonumber
R_{\mu\nu} - \frac{1}{2}\,g_{\mu\nu}R &=& \Lambda g_{\mu\nu}
- \gamma_2\hbar\Big(RR_{\mu\nu} + g_{\mu\nu}\Box R - R_{;\mu\nu} - \frac{1}{4}\,g_{\mu\nu}R^2\Big)
- \frac{\gamma_3\hbar}{a^4(\bar{\eta})}\left(g_{\mu\nu} - \frac{4}{3}\,g_{i\mu}\,\delta_{i\nu}\right) 
\\[1.5mm]&&
- \gamma_1\hbar\left(\frac{2}{3}\,RR_{\mu\nu} - R_{\mu\lambda}R_{\nu}^{\lambda} - \frac{1}{4}\,g_{\mu\nu}R^2 
+ \frac{1}{2}\,g_{\mu\nu}R_{\lambda\rho}R^{\lambda\rho}\right)
+ \text{O}(\hbar^2)\,,
\eeqa
where $\langle \hat{T}_{\mu\nu} \rangle_\text{O}$ has been already substituted, 
$\gamma_1 \equiv 1/360\pi$, $\gamma_2$ is a finite coefficient 
depending on the mass scale $\mu$, and $\gamma_3 \equiv 1/60\pi$.

Using the order reduction method for throwing away spurious solutions~\cite{Parker&Simon}, one obtains the following first order nonlinear 
equation
\beqa
3(a^{\prime})^2 - 3a^2 &=& \Lambda a^4\left(1+ \frac{\Lambda}{3}\,\gamma_1\hbar\right) - \gamma_3\hbar + \text{O}(\hbar^2)\,. 
\eeqa
where prime stands for the differentiation over the conformal time $\bar{\eta}$. It is solved up to the second order of $\hbar$ by
\beqa\label{eq:ads4:corrected-a}
a(\bar{\eta}) &=& \frac{1}{\widetilde{H}\cosh(\bar{\eta})} + \frac{\gamma_3\hbar}{12}\sqrt{\frac{-\Lambda}{3}}
\left(\cosh(\bar{\eta}) + 3\,\frac{\bar{\eta}\sinh(\bar{\eta}) - \cosh(\bar{\eta})}{\cosh^2(\bar{\eta})}\right) +  \text{O}(\hbar^2)\,,
\eeqa
where $3\widetilde{H}^2 = -\Lambda (1 + \gamma_1\hbar\Lambda/3)$, an integration constant has been fixed by the condition: 
$a(0) = 1/\widetilde{H} + \text{O}(\hbar)$, and a solution corresponding to the shift in time has been omitted. 

From the equation \eqref{eq:ads4:corrected-a} one immediately sees that there exists $\bar{\eta}_0 > 0$, such that for 
$|\bar{\eta}| > \bar{\eta}_0$
the correction of order $\hbar$ in $a(\bar{\eta})$ starts to dominate.
It means that anti de Sitter space with the quantum field being in $|0_\text{O}\rangle$ is unstable, provided it has been 
somehow realized. 

If the field is in $|0_\text{S}\rangle$, then the quantum corrections merely lead to a tiny decrease of the curvature.

\section{Massive scalar field in AdS$_2$}
\label{sec:ads2}

Two-dimensional anti de Sitter space can be obtained from  \eqref{ads4:hyperboloid} by setting $x_2 = x_3 = 0$.
The metric tensor in static or open coordinates are obtained by taking $\theta = 0$ and $\phi = \text{const}$ in 
\eqref{app:coordinates:metric:global:conformal} or \eqref{eq:app:conformal-frw-like}, where
$\chi \in (-\frac{\pi}{2},+\frac{\pi}{2})$ and $\bar{\chi} \in \mathbf{R}$, respectively.

\paragraph*{Static AdS$_2$} 

The massive quantum scalar field can be expanded as
\beqa
\hat{\phi}(x) &=& 
\sum\limits_{n=0}^{+\infty}
\left\{
\begin{array}{l}
u_{n}^{\phantom{\dagger}}(x)\hat{a}_{n}^{\phantom{}} + u_{n}^*(x)\hat{a}_{n}^{\dagger}\,,
\\[2.5mm]
v_{n}^{\phantom{\dagger}}(x)\hat{b}_{n}^{\phantom{}} + v_{n}^*(x)\hat{b}_{n}^{\dagger}\,.
\end{array}
\right.
\eeqa
where the positive-frequency modes with respect to the translation operator along time coordinate are
given by
\bsubeqs
\beqa
u_n(x) &=& A_n\,e^{-i(2n+1+\lambda)\eta}\,(\cos\chi)^{\lambda}\,C_{2n+1}^{\lambda}(\sin\chi)\,,
\\[2mm]
v_n(x) &=& B_n\,e^{-i(2n+\lambda)\eta}\,(\cos\chi)^{\lambda}\,C_{2n}^{\lambda}(\sin\chi)\,,
\eeqa
\esubeqs
$\lambda^2 - \lambda = 2\xi_\text{R} + (m/H)^2$ and $\lambda > 1/2$, where $\xi_\text{R}$ is the coupling constant of the field with
the scalar curvature, $A_n$ and $B_n$ are real normalization coefficients. The energy flux through the spatial boundary of anti de Sitter space vanishes for both sets of modes, thus the 
energy of the field is conserved. In addition, the scalar product is preserved as well~\cite{Sakai&Tanii}. 
The static vacuum is defined as a state being annihilated by both $\hat{a}_{n}$ and $\hat{b}_{n}$ for any $n \in \mathbb{N}_0$.

\paragraph*{Open AdS$_2$}

The metric tensor in open coordinates is homogeneous, therefore the field can be expanded through the eigenfunctions
of the translation operator along the spatial coordinate
\beqa
\hat{\phi}(\bar{x}) &=& \int\limits_{-\infty}^{+\infty}dp\Big(w_p(\bar{x})\,\hat{c}_{p}^{\phantom{}}
+ w_p^*(\bar{x})\,\hat{c}_{p}^{\dagger}\Big)\,,
\eeqa
where $w_p(\bar{x}) = \varphi_p(\bar{\eta})\,e^{ip\bar{\chi}}$. The function $\varphi_k(\bar{\eta})$ satisfies
\beqa\label{eq:milne-ads:modes}
\Big(\partial_{\bar{\eta}}^2 +p^2 + a^2(\bar{\eta})\big[m^2+\xi_\text{R}R(\bar{\eta})\big]\Big)\varphi_{p}(\bar{\eta}) &=& 0\,,
\eeqa
where $R(\bar{\eta})$ is the Ricci scalar.

The general solution of \eqref{eq:milne-ads:modes} with $a(\bar{\eta}) = (H\cosh\bar{\eta})^{-1}$ can be written in terms 
of Legendre Q-functions~\cite{Gradshteyn&Ryzhik} in the following form 
\beqa\label{eq:milne-ads:ads}
\varphi_{p}(\bar{\eta}) &=&C_p(\cosh\bar{\eta})^{\frac{1}{2}}
\Big(Q_{\nu}^{\mu}(i\sinh\bar{\eta}) + \gamma_pQ_{-\nu -1}^{\mu}(i\sinh\bar{\eta})\Big)\,,
\eeqa
where $\nu = i\omega-\frac{1}{2}$, $\omega = |p|$, $\mu = \lambda - \frac{1}{2}$,
$C_p$ and $\gamma_p$ are integration constants. The associated Legendre functions are defined in the complex 
plane except on $(-\infty,1)$. In particular, it implies the discontinuity~\cite{Gradshteyn&Ryzhik}:  
$Q_{\nu}^{\mu}(+i0) \neq Q_{\nu}^{\mu}(-i0)$. Therefore, 
we take the solution \eqref{eq:milne-ads:ads} for $\bar{\eta} > 0$ and perform analytical continuation on negative time: 
$Q_{\nu}^{\mu}(i\sinh\bar{\eta}) \rightarrow e^{i\pi\mu}(Q_{\nu}^{\mu}(i\sinh\bar{\eta}) - i\pi P_{\nu}^{\mu}(i\sinh\bar{\eta}))$, 
where $\bar{\eta} < 0$ on the right-hand side.

\subsection{Bogolyubov transformation}

The choice of $\gamma_p$ determines no-particle state in open coordinates. To compare them with static vacuum, one 
needs to compute the Bogolyubov coefficients. In order to simplify calculations, one may choose a surface $\bar{\eta} = +0$. 
The same result must be obtained in the case $\bar{\eta} = - 0$ after performing the analytic continuation. In both cases we do 
get the same result
\beqa\label{eq:milne-ads:ratio}
\frac{\beta_{pn}}{\alpha_{pn}} &=& \mp\frac{(1+\gamma_p)e^{-2\pi i\lambda}}{\gamma_pe^{-\pi\omega}+e^{\pi\omega}}\,,
\eeqa
where the minus sign is taken for $u$-modes and the plus sign is for $v$-modes.

According to \eqref{eq:milne-ads:ratio}, the modes \eqref{eq:milne-ads:ads} with $\gamma_p = -1$ corresponds to static vacuum. Thus, a comoving observer in open AdS will see no particles along his movement in this case.

In the limits $\bar{\eta} \rightarrow \pm\infty$, the scale factor vanishes and $a^{\prime}/a \rightarrow \mp 1$. The zeroth-order 
approximation of the WKB-type solution~\cite{Birrell&Davies} gives
\beqa
\varphi_p(\bar{\eta}) &\sim& e^{-i\omega\bar{\eta}}\,, \quad \bar{\eta} \;\;\rightarrow\;\; \pm\infty\,.
\eeqa
The next order gives a vanishing contribution at past and future time infinities. The exact solution will coincide with the
approximate WKB solution in remote past if one chooses 
$\gamma_p = -e^{\pi\omega}\sin(\pi\lambda)\csc(\pi(\lambda+i\omega))$. Unless $\lambda$ is a positive integer, out-vacuum
is filled by particles defined with respect to in-vacuum. However, such vacuum would be unphysical, because it implies 
a discontinuity in a density number of particles in the points between the wedges.
For $\lambda \in \mathbb{N}$\footnote{In the conformal case, the zero-mode solution has to be taken into account in static AdS. However, one can directly show that this does not change our conclusions.}, in-vacuum and out-vacuum coincide and equal to the conformal one at the conformal time infinities. Plugging $\gamma_p = 0$ into \eqref{eq:milne-ads:ratio}, one finds $|\beta_{pn}/\alpha_{pn}| = e^{-\pi\omega}$. Thus, 
if the field occupies this vacuum, then a freely-moving observer will register a thermal bath.

\subsection{Relation between $|0_\text{S}\rangle$ and Minkowski vacuum $|0_\text{M}\rangle$}

One can show that for open vacuum just defined the renormalized stress tensor $\langle T_{\nu}^{\mu} \rangle_\text{O}$  diverges 
at the horizons as it does in the four-dimensional case.

Assume $\lambda = 1$, i.e. the conformal invariant case, and the scale factor behaves itself with the conformal time according to
\beqa
a(\bar{\eta}) &=& a_0 + \frac{1}{H\cosh\bar{\eta}}\,, \quad a_0 = \text{const} > 0\,.
\eeqa
The universe becomes Minkowski spacetime at past and future time infinities and anti de Sitter spacetime for 
$a_0H\cosh\bar{\eta} \ll 1$. 
At remote past it is natural to define Minkowski vacuum, such that $\phi_p(\bar{\eta}) \sim e^{-i\omega\bar{\eta}}$, where 
$\omega = |p|$ as before. 
The AdS modes which correspond to the positive-frequency modes in Minkowski space at $\bar{\eta} \rightarrow \pm\infty$ are specified 
by setting $\gamma_p = 0$ in \eqref{eq:milne-ads:ads}. 
Thus, a geodesic observer that starts his journey from Minkowski spacetime will detect a thermal bath 
of particles at the AdS stage of the universe evolution. Note that the renormalized stress tensor 
vanishes at past and future infinities.

\section{Concluding remarks}

A quantization of the conformal scalar field in four-dimensional AdS in static and open coordinates leads to a 
definition of inequivalent quantum vacua. The former is associated with a freely-moving observer being located at the
origin. The latter is a conformal 
vacuum associated with open coordinate system that resembles FRW-like universe with the negative spatial curvature and the
scale factor periodically expanding from and contracting to the zero value: $a(\bar{t}) = H^{-1}\cos(H\bar{t})$. Under the assumption that the quantum field occupies the conformal vacuum, $|0_\text{O}\rangle$, it has been shown that a geodesically moving observer will register a thermal bath of particles. The temperature of this thermal condensate redshifts as it does for the ordinary thermal radiation:
$T_\text{AdS} = (2\pi a(\bar{t}))^{-1}$. This implies that at the moments of time 
$H\bar{t} = \frac{\pi}{2} + \pi k$, where $k \in \mathbb{N}$ , the anti de Sitter temperature diverges. Physically, this means that
the conformal vacuum is cosmologically unstable at the quantum level, provided it has be somehow prepared. Thus, static vacuum $|0_\text{S}\rangle$ is a better candidate for the true vacuum in AdS and, as a result, a geodesic observer will not register particles during his movement. 

However, one may imagine a universe that looks like open static universe (OSU) in remote past and future. In this case, the scale factor approaches a nonzero constant $a_0$ in the limits $\bar{\eta} \rightarrow \pm\infty$. Since OSU is static, it admits a global time-like
Killing vector, such that a definition of a particle in OSU is as natural as in Minkowski spacetime~\cite{Birrell&Davies}. Thus, 
a geodesic observer will register a thermal radiation at the AdS stage of the universe evolution when $a_0H\cosh\bar{\eta} \ll 1$.

The non-conformal scalar field has been examined in two-dimensional AdS as well. We have found similar results, provided
the field is in open vacuum that has been defined as being equal to the conformal one at remote past and future for $\lambda \in \mathbb{N}$. 
In the conformal case, we have considered a universe that is Minkowski space at the time infinities and has
the AdS stage in between. A geodesic observer in such a universe will register a thermal radiation when the universe
becomes quasi anti de Sitter spacetime.

\section*{
ACKNOWLEDGMENTS}

I am thankful to Prof. Mukhanov for the suggestion to consider this problem as well as
for the valuable discussions during the research.

\begin{appendix}
\section{Coordinate parameterizations of AdS}
\label{app:coordinates}

\paragraph*{Static or global coordinates}
\beqa\nonumber
x_0 &=& (H^{-2} + r^2)^{\frac{1}{2}}\cos(Ht)\,, 
\\[1mm]\nonumber
x_1 &=& r\cos\theta\,,
\\[1mm]
x_2 &=& r\sin\theta\cos\varphi\,,
\\[1mm]\nonumber
x_3 &=& r\sin\theta\sin\varphi\,,
\\[1mm]\nonumber
x_4 &=& (H^{-2} + r^2)^{\frac{1}{2}}\sin(Ht)\,,
\eeqa
where $t \in (-\infty,+\infty)$, $r \in [0,+\infty)$, $\theta \in [0,\pi]$ and $\varphi \in [0,2\pi)$. The line element in these coordinates 
is given by
\beqa\label{app:coordinates:metric:global}
ds^2 &=& \big(1 + H^2r^2\big)dt^2 -\frac{dr^2}{1+H^2r^2} - r^2d\Omega^2\,.
\eeqa

Introducing new variables $Ht = \eta$ and $Hr = \tan\chi$, \eqref{app:coordinates:metric:global} becomes
\beqa\label{app:coordinates:metric:global:conformal}
ds^2 &=& \frac{1}{H^2\cos^2\chi}\Big(d\eta^2 - d\chi^2 - \sin^2\chi\, d\Omega^2\Big)\,.
\eeqa

\paragraph*{FRW-like or open coordinates}
\beqa\nonumber
x_0 &=& -H^{-1}\sin(H\bar{t})\,, 
\\[1mm]\nonumber
x_1 &=& H^{-1}\cos(H\bar{t})\sinh\bar{r}\cos\bar{\theta}\,,
\\[1mm]
x_2 &=& H^{-1}\cos(H\bar{t})\sinh\bar{r}\sin\bar{\theta}\cos\bar{\varphi}\,,
\\[1mm]\nonumber
x_3 &=& H^{-1}\cos(H\bar{t})\sinh\bar{r}\sin\bar{\theta}\sin\bar{\varphi}\,,
\\[1mm]\nonumber
x_4 &=& H^{-1}\cos(H\bar{t})\cosh\bar{r}\,,
\eeqa
where $\bar{t} \in (-\infty,+\infty)$, $\bar{r} \in [0,+\infty)$, $\bar{\theta} \in [0,\pi]$ and $\bar{\varphi} \in [0,2\pi)$, such those
\beqa\label{app:coordinates:metric:synchronous}
ds^2 &=& d\bar{t}^2 - a^2(\bar{t})\Big(d\bar{r}^2 + \sinh^2\bar{r}\, d\bar{\Omega}^2\Big)\,,
\quad a(\bar{t}) \;\;=\;\; H^{-1}\cos(H\bar{t})\,.
\eeqa

Introducing new variables $\sin(H\bar{t}) = \pm\tanh(\bar{\eta})$, where the plus sign is taken for the upper wedge and
the minus sign is for the lower wedge (see Fig. 1), and $\bar{r} = \bar{\chi}$, 
\eqref{app:coordinates:metric:synchronous} takes the following conformal form
\beqa\label{eq:app:conformal-frw-like}
ds^2 &=& a^2(\bar{\eta})\Big(d\bar{\eta}^2 - d\bar{\chi}^2 - \sinh^2(\bar{\chi})d\bar{\Omega}^2\Big)\,, \quad
a(\bar{\eta}) \;\;=\;\; \frac{1}{H\cosh(\bar{\eta})}\,.
\eeqa

\section{Evaluation of integrals}
\label{app:integrals}

\begin{figure}
\includegraphics[width=7.5cm]{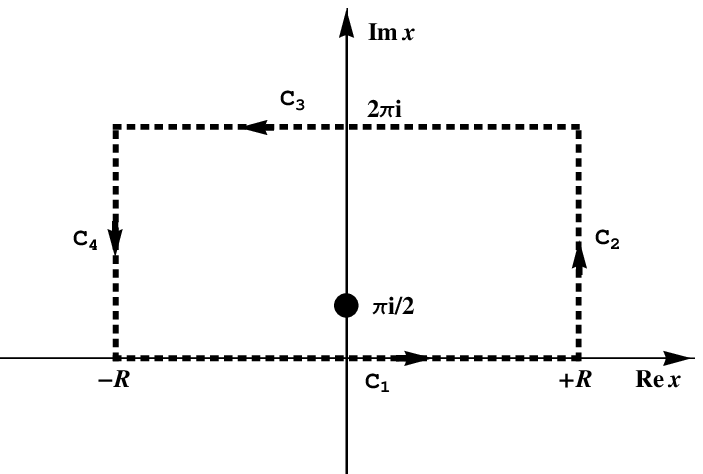}
\includegraphics[width=7.5cm]{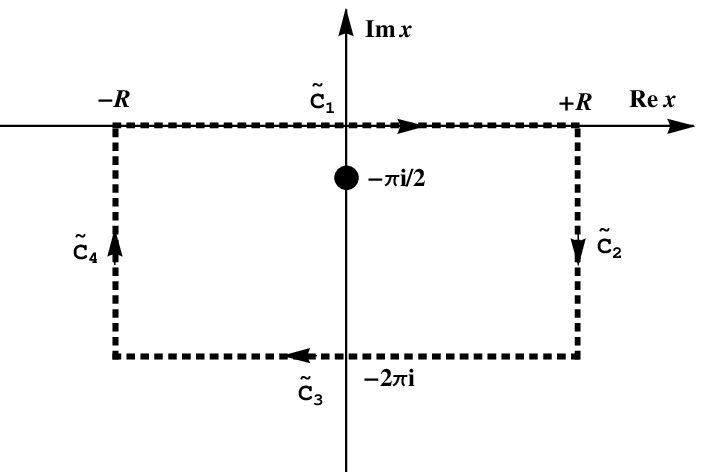}
\caption{Contours used in the residue theorem to evaluate integrals $J_n^{\pm}(\alpha)$ and $I_n^{\pm}(\alpha)$.}
\label{fig:contours}
\end{figure}

Let us consider the following regularised integral
\beqa
J_{n}^{+}(\alpha) &=& \int\limits_{-\infty}^{+\infty}dx\,e^{i\alpha x - \varepsilon x^2}\left(\frac{e^{x} + i}{e^{x} -i}\right)^n\,,
\;\; \text{where} \;\; \alpha > 0\,, \;\; n \ge 1 \;\; \text{and} \;\;  \varepsilon \rightarrow +0\,.
\eeqa
The integrand has poles at imaginary values of $x$. One of them is at $x = \pi i/2$. The residue theorem applied for the 
contour $C = C_1\cup C_2\cup C_3\cup C_4$ plotted in Fig.~\ref{fig:contours} in the limit $R \rightarrow +\infty$ gives
\beqa
J_{n}^{+}(\alpha) &=& \frac{\pi i\exp(\pi\alpha)}{\sinh(\pi\alpha)}\,
\underset{\pi i/2}{\textrm{res}}
\left(e^{i\alpha x}\left(\frac{e^{x} + i}{e^{x} -i}\right)^n\right)
\\[1.5mm]\nonumber
&=& (-1)^{n+1}
\frac{2\pi i\exp\left(\frac{\pi\alpha}{2}\right)}{\sinh\left(\pi\alpha\right)}\,
\frac{\Gamma\left(n - i\alpha\right)}{\Gamma\big(n\big)\Gamma\left(1 - i\alpha\right)}\;
_2F_1\left(1 - n,1 +i\alpha, 1 - n + i\alpha; -1\right)\,.
\eeqa
where the regularization parameter has been suppressed.

Analogously, for the integral
\beqa
J_{n}^{-}(\alpha) &=& \int\limits_{-\infty}^{+\infty}dx\,e^{i\alpha x - \varepsilon x^2}\left(\frac{e^{x} - i}{e^{x} + i}\right)^n\,,
\;\; \text{where} \;\; \alpha > 0\,, \;\; n \ge 1 \;\; \text{and} \;\;  \varepsilon \rightarrow +0\,
\eeqa
along the contour $\tilde{C}$  in the limit $R \rightarrow +\infty$ (see Fig.~\ref{fig:contours}) we find
\beqa
J_{n}^{-}(\alpha) &=& \frac{\pi i\exp(-\pi\alpha)}{\sinh(\pi\alpha)}\,
\underset{-\pi i/2}{\textrm{res}}
\left(e^{i\alpha x}\left(\frac{e^{x} - i}{e^{x} + i}\right)^n\right)
\\[1.5mm]\nonumber
&=& (-1)^{n+1}
\frac{2\pi i\exp\left(-\frac{\pi\alpha}{2}\right)}{\sinh\left(\pi\alpha\right)}\,
\frac{\Gamma\left(n - i\alpha\right)}{\Gamma\big(n\big)\Gamma\left(1 - i\alpha\right)}\;
_2F_1\left(1 - n,1 +i\alpha, 1 - n + i\alpha; -1\right)\,.
\eeqa

Repeating similar calculations, we further obtain for $n \ge 0$
\beqa\nonumber
I_{n}^{+}(\alpha) &=&  \int\limits_{-\infty}^{+\infty}dx\,e^{(i\alpha+1)x}\frac{\big(e^{x} + i\big)^{1+2n}}{\big(e^{x} -i\big)^{3+2n}}
\\[1.5mm]
&=& +\frac{i\pi^2\exp\left(\frac{\pi\alpha}{2}\right)}{2\sinh^2(\pi\alpha)}\,
\frac{_2F_1\left(-2 - 2n,i\alpha, -1 - 2n + i\alpha; -1\right)}{\Gamma\left(3+2n\right)\Gamma\left(-i\alpha\right)\Gamma\left(i\alpha-1-2n\right)}\,
\eeqa
and
\beqa\nonumber
I_{n}^{-}(\alpha) &=&  \int\limits_{-\infty}^{+\infty}dx\,e^{(i\alpha+1)x}\frac{\big(e^{x} - i\big)^{1+2n}}{\big(e^{x} + i\big)^{3+2n}}
\\[1.5mm]
&=& -\frac{i\pi^2\exp\left(-\frac{\pi\alpha}{2}\right)}{2\sinh^2(\pi\alpha)}\,
\frac{_2F_1\left(-2 - 2n,i\alpha, -1 - 2n + i\alpha; -1\right)}{\Gamma\left(3+2n\right)\Gamma\left(-i\alpha\right)\Gamma\left(i\alpha-1-2n\right)}\,.
\eeqa

\section{Particular properties of hypergeometric function ${}_2F_1\left(\alpha, \beta; \gamma; z\right)$ and derivation of $M_{n,l+1}^{\pm}$
as a sum of $M_{n,l}^{\pm}$ and $M_{n+1,l}^{\pm}$}
\label{app:property}

To derive \eqref{eq:ads4:mp}, one utilizes the following properties of the hypergeometric function~\cite{Gradshteyn&Ryzhik}:
\beqa\label{eq:app:properties:first}
{}_2F_1\left(\alpha, \beta; \gamma + 1; z\right) &=&
\frac{\gamma}{(\gamma-\alpha)(\gamma-\beta)}
\\[1mm]&&\nonumber
\times\;\left((1-z)\frac{d}{dz}\,{}_2F_1\left(\alpha, \beta; \gamma; z\right)
+(\gamma-\alpha-\beta){}_2F_1\left(\alpha, \beta; \gamma; z\right)\right)\,.
\eeqa
\beqa\label{eq:app:properties:second}\nonumber
\frac{d}{dz}\,{}_2F_1\left(\alpha, \beta; \gamma; z\right) &=&
\frac{1}{(\alpha-\beta)z(1-z)^2}
\,\Big((\alpha-\gamma)(\beta-\gamma+\alpha z){}_2F_1\left(\alpha-1; \beta, \gamma; z\right)
\\[1mm]&&
-(\beta-\gamma)(\alpha-\gamma+\beta z){}_2F_1\left(\alpha, \beta-1; \gamma; z\right)\Big)\,.
\eeqa
\beqa\label{eq:app:properties:third}
{}_2F_1\left(\alpha, \beta+1; \gamma; z\right) &=&
\frac{1}{(\alpha-\beta-1)(1-z)}
\\[1mm]&&\nonumber
\times\;\Big((\alpha-\gamma){}_2F_1\left(\alpha-1, \beta+1; \gamma; z\right)
-(\beta-\gamma+1){}_2F_1\left(\alpha, \beta; \gamma; z\right)\Big)\,.
\eeqa

The derivation of \eqref{eq:ads4:mp} consists in the three steps. First,
having taken $l$ exchanged into $l+1$ in \eqref{eq:ads4:m}, one uses \eqref{eq:app:properties:first} to 
express ${}_2F_1(\frac{1}{2}-i\alpha,\frac{1}{2}+i\alpha;l+\frac{5}{2}; z)$
through ${}_2F_1(\frac{1}{2}-i\alpha,\frac{1}{2}+i\alpha;l+\frac{3}{2}; z)$ and its first derivative over $z$.
Second, after integration by parts, one then employs \eqref{eq:app:properties:second} to rewrite 
$\frac{d}{dy}{}_2F_1(-n,n+l+3;\frac{3}{2}; y)$ via ${}_2F_1(-1-n,n+l+3;\frac{3}{2}; y)$ and ${}_2F_1(-n,n+l+2;\frac{3}{2}; y)$.
Third, the equality \eqref{eq:app:properties:third} is applied in order to rewrite ${}_2F_1(-n,n+l+3;\frac{3}{2}; y)$ as 
a sum of ${}_2F_1(-1-n,n+l+3;\frac{3}{2}; y)$ and ${}_2F_1(-n,n+l+2;\frac{3}{2}; y)$. This sequence of computations leads
to the formula \eqref{eq:ads4:mp}.

\section{Commutation relations}
\label{app:relations}

The Bogolyubov coefficients satisfy normalization condition~\cite{Birrell&Davies,Mukhanov&Winitzki}. 
Employing \eqref{eq:ads4:ratio}, one obtains
\beqa\label{eq:app:relations:1}
\sum\limits_{nlm}\alpha_{\omega^{\prime}l^{\prime}m^{\prime},nlm}\,\alpha_{\omega^{\prime\prime}l^{\prime\prime}m^{\prime\prime},nlm}^*
&=& \frac{\exp\left(\pi\omega^{\prime}\right)}{2\sinh\left(\pi\omega^{\prime}\right)}\,
\delta(\omega^{\prime} - \omega^{\prime\prime})\,\delta_{l^{\prime}l^{\prime\prime}}\,\delta_{m^{\prime}m^{\prime\prime}}\,.
\eeqa
Hence, considering a commutator of $\hat{d}_{\omega lm}$ and its hermitian conjugate, one derives 
\eqref{eq:ads4:commutations1}, provided $\tanh\theta_{\omega} = e^{-\pi\omega}$.

The first commutator in \eqref{eq:ads4:commutations1} vanishes, because $\hat{a}_{\omega lm}^{\phantom{}}$ and $\hat{a}_{\omega lm}^\dagger$ commute. To show that the second commutator in \eqref{eq:ads4:commutations1} vanishes as well, one 
has to note that it follows from \eqref{eq:app:relations:1}
\beqa
\sum\limits_{n}N_{nl}(\omega^{\prime})N_{nl}(-\omega^{\prime\prime})M_{n,l}^{+}(\omega^{\prime})M_{n,l}^{-}(-\omega^{\prime\prime})
&\propto& \delta(\omega^{\prime} - \omega^{\prime\prime})\,
\eeqa
is valid, $(M_{n,l}^{+}(\omega))^* = - M_{n,l}^{-}(-\omega)$ and $N_{nl}^*(-\omega) = i^{2l+2}N_{nl}(\omega)$ have been used, where 
\beqa
N_{nl}(\omega) &=& \frac{(-1)^{5/4}2^{l+1}i^l\Gamma\left(n+l+2\right)\Gamma\left(1+l+i\omega\right)}
{\sqrt{\pi}\Gamma\left(1+n\right)\Gamma\left(1+i\omega\right)\Gamma\left(l+3/2\right)}
\left(\frac{\omega\Gamma\left(2+2n\right)}{\Gamma\left(3+2l+2n\right)}\right)^{\frac{1}{2}}.
\eeqa
Now from
\beqa\nonumber
\big[\hat{d}_{\omega^{\prime} l^{\prime}m^{\prime}}^{\phantom{}},\hat{f}_{-\omega^{\prime\prime}l^{\prime\prime}m^{\prime\prime}}^{\dagger}\big] 
&\propto& \sum\limits_{nlm}\alpha_{\omega^{\prime}l^{\prime}m^{\prime},nlm}\,\beta_{-\omega^{\prime\prime}l^{\prime\prime}m^{\prime\prime},nlm}
\\[1.5mm]
&\propto& \sum\limits_{n}N_{nl}(\omega^{\prime})N_{nl}(-\omega^{\prime\prime})M_{n,l}^{+}(\omega^{\prime})M_{n,l}^{-}(-\omega^{\prime\prime})
\;\;\propto\;\; \delta(\omega^{\prime} - \omega^{\prime\prime})\,,
\eeqa
one finds
$
\big[\hat{d}_{\omega^{\prime} l^{\prime}m^{\prime}}^{\phantom{}},\hat{f}_{\omega^{\prime\prime}l^{\prime\prime}m^{\prime\prime}}^{\dagger}\big] \propto \delta(\omega^{\prime} + \omega^{\prime\prime}) = 0
$.
\end{appendix}


\end{document}